\documentclass[onecolumn,submit]{scias} 

\FrontMatterExtraVskip{0.0pt} 

\usepackage{caption}
\usepackage{amsmath}
\usepackage{algorithm}
\usepackage{algorithmic}
\usepackage{subfigure}

\title{A Combination Model Based on Sequential General Variational Mode Decomposition Method for Time Series Prediction }

\author[*]{Wei Chen}
\author{Yuanyuan Yang}
\author{Jianyu Liu}
\author[*]{Yong Zhang}
\author{Zhenfeng Chen}
\author{Huaicheng Liu}
\author{Jinwei Hu}
\author{Bin Wan}
\address{JiangXi University of Finances and Economics}

\ead{chenwei@jxufe.edu.cn} 

\abstract{Accurate prediction of financial time series is a key concern for market economy makers and investors. The article selects online store sales and Australian beer sales as representatives of non-stationary, trending, and seasonal financial time series, and constructs a new SGVMD-ARIMA combination model in a non-linear combination way to predict financial time series. The ARIMA model, LSTM model, and other classic decomposition prediction models are used as control models to compare the accuracy of different models. The empirical results indicate that the constructed combination prediction model has universal advantages over the single prediction model and linear combination prediction model of the control group. Within the prediction interval, our proposed combination model has improved advantages over traditional decomposition prediction control group models.}


\MSC{  }

\begin{document}
\frontmatter

\section{Introduction}
Financial time series forecasting is an important method for studying financial market volatility\cite{databased, databased1}. By using historical data generated by financial markets to establish predictive models\cite{modelbased}, we can explore inherent volatility patterns and predict future trends\cite{bigdata, bigdata1}. Financial time series, as a special type of time series\cite{statisticalin,stachara}, have commonalities with time series, such as obtaining noise that may be present in the data\cite{market, market1}. It also has the particularity of financial data, such as the nonlinear chaotic relationship between data, making it more difficult to grasp its fluctuation patterns\cite{power, power1}. Therefore, how to accurately reveal the trend of changes in financial time series and make reasonable predictions of financial time series is a key concern for the academic community and practitioners\cite{stock, stock1}.

The research objects of financial time series forecasting are diverse\cite{precipitation, precipitation1, economic, engineering}, and the research methods have also undergone extensive evolution\cite{naturalsci}, from statistical models to intelligent models, from single models to composite models. Statistical models\cite{tech}, represented by the Differential Autoregressive Moving Average (ARIMA) model and the Holt-Winters forecasting method model\cite{arma, arma1, arima, arima1, HW}, have a wide range of applications. For example, combining ARIMA with various decomposition algorithms such as Empirical Mode Decomposition (EMD) and Variational Mode Decomposition (VMD) for predicting complex time series; For example, using an improved ARMA model for stock market forecasting. However, the above models need to be built on the basis of stable sequence data, and usually require testing and preprocessing of the original data, which may lead to the loss of some hidden information, especially in big data samples, and this disadvantage is easily magnified.

With the development of computer technology, intelligent models represented by artificial neural networks (ANNs) are gradually emerging. This type of model is good at handling incomplete, fuzzy, uncertain, or irregular data, and has a good fit to nonlinear relationships. Shallow neural networks represented by backpropagation neural networks (BPNN) and shallow machine learning represented by support vector machines (SVM) are also widely used in financial market prediction. However, shallow neural networks do not consider the temporal nature of data, and financial time series often have certain long-term dependencies. Therefore, recurrent neural networks (RNNs) with memory function have become the latest choice. The output of RNN at a certain moment can be used as input to feedback to neurons again, and this cascade structure is very suitable for time series data, which can preserve the dependency relationships in the data. However, in a large amount of practice, it has been found that standard RNNs are difficult to achieve long-term information preservation, and may be plagued by vanishing or exploding gradients during training. Therefore, the Long Short Term Memory Neural Network (LSTM) used to improve the standard RNN is more efficient in practical applications of financial time series prediction\cite{LSTM}.

Regardless of the individual model, it has its own advantages and limitations, so we will propose an improved VMD model. The formation of this method can be divided into two steps: (1) introducing the generalized form of the Variational Mode Decomposition (GVMD) ; (2) proposing the Sequentialized and generalized Variational Mode Decomposition (SGVMD) for the case of unknown modal number\cite{SGVMD}. These two algorithms are crucial for time series decomposition. In the decomposition step of combination model, the extracted components are predicted. For the trend component, an ARIMA model is directly constructed for prediction. For the narrowband FM-AM component , the upper and lower envelope lines are first extracted, and then separate ARIMA models are constructed for each envelope line for prediction\cite{autoreg, autoreg1}. The operation of time series that we propose, called SGVMD-E-ARIMA model because of the envelop prediction process among\cite{movinaver, nonstaion}. We also could predict every component directly, called SGVMD-ARIMA model. Decomposition models can effectively explore the implicit linear relationships in data, while adding constraints has its own uniqueness in handling nonlinear relationships. However, the large amount of data generated in financial markets is difficult to determine whether it is a completely linear or nonlinear relationship, so it is necessary to combine the advantages of decomposition models and prediction models to construct a combination model\cite{defferencing, season}. The common approach to constructing composite models is to decompose the data, fit the linear and nonlinear parts separately through a constrained decomposition model and a prediction model, and then superimpose them to obtain the prediction results. Based on this idea, we built a combination model to predict the financial time series, and confirmed that the combination model has advantages over a single model\cite{nonlinear, modeltrain}.

The above research has laid the foundation for the selection of research content and methods in this article. However, the existing research mostly focuses on a certain financial time series, such as sales, stock index, exchange rate, etc. It is difficult to prove whether the same model is universal for different financial time series\cite{triesmooth}. In addition, existing combination models are often combined in a linear manner\cite{ewma, ewma1}, that is, the decomposition model part and the prediction model part are directly coupled. But if there is no direct coupling relationship between the two parts, this combination method is likely to lead to a decrease in predictive performance\cite{rnn}. Therefore, this article constructs a decomposition combination prediction method with added constraints, namely the SGVMD-ARIMA model, aiming to combine the advantages of the two models to predict various financial time series such as sales, and study the prediction accuracy and universality of this combination model in different financial time series\cite{emd, emd1}.

\section{Methodology}
\subsection{Variational Mode Decomposition (VMD)}

The Variational Mode Decomposition (VMD) algorithm, a powerful signal processing technique, has gained prominence in extracting inherent modes from complex signals\cite{vmd, vmd1}. VMD employs a variational optimization framework to decompose a signal into a set of adaptive modes characterized by their respective center frequencies and amplitudes\cite{wu}. The method is particularly adept at handling non-stationary and nonlinear signals, making it valuable in various applications, such as biomedical signal processing and communication systems\cite{imf, envelope}.

Unlike Huang's concept of Intrinsic Modal Component (IMF), the VMD algorithm redefines the intrinsic modal function of a finite bandwidth with stricter constraints. This intrinsic modal component is defined as the component modal function of amplitude modulation and frequency modulation, expressed mathematically as
\begin{equation} 
	\begin{split}
		u_k(t) = A_k(t)cos(\phi_k(t))
	\end{split}
\end{equation}
Where $A_k(t)$ is the signal $u_k(t)$'s envelope amplitude, and $\phi_k(t)$ is the instantaneous phase. It should be noted that the components represented by this function also meet the constraints of EMD.

The VMD method is a non-recursive variational signal decomposition method, and its overall framework is a variational problem. The VMD algorithm is founded on the minimization of a cost function, typically defined as follows:
\begin{equation} 
	\begin{split}
		\mathop{\min}_{\{u_k\},\{\omega_k\}}
 \{ \sum_k  \lVert \partial_t [(\delta(t) + \frac{j}{\pi t}) * u_k(t)] e^{-j\omega_k t} \rVert_2^2 \\
		s.t. \quad \sum_k u_k = f
	\end{split}
\end{equation}
The above equation represents the modal function $u_k$ and center frequency $\omega_k$ when the bandwidth sum of the center frequencies of each modal component is minimized.

The Variable Mode Decomposition (VMD) algorithm has some drawbacks. Firstly, it may exhibit sensitivity to initial parameters, making its performance dependent on careful parameter tuning. Additionally, VMD may face challenges in handling non-stationary signals and complex data structures due to its reliance on a predefined number of modes. Despite its effectiveness in certain applications, these limitations highlight the need for further refinement and adaptation to diverse data scenarios.

\subsection{General Variational Mode Decomposition (GVMD) and Sequential General Variational Mode Decomposition (SGVMD)}
In order to extend the applicability of the VMD algorithm, the GVMD algorithm is considered. Essentially, it is a generalization and eigenvalueization of the VMD algorithm, which extends the variational mode decomposition method to the general field. The general form of the loss function of the algorithm is shown in equation (1):
\begin{equation} 
	\begin{split}
		\mathcal{L}\left(\left\{\hat{u}_{i}\right\}\right)=\left\|\hat{f}-\sum_{i=1}^{K} \hat{u}_{i}\right\|^{2}+\alpha \sum_{i=1}^{K}\left\|g\left(\hat{u}_{i}\right)\right\|^{2} 
	\end{split}
\end{equation}

Where the function $g\left(\hat{u}_{i}\right)$ represents the characteristic feature of ${\hat{u}}_{i}$.

VMD is just a special case of GVMD when the components have narrowband characteristics in the frequency domain. GVMD is more general and not limited to the time and frequency domains, nor does it have fixed requirements for component characteristics.

However, both VMD and GVMD algorithms have a common drawback, which is the need to know the modal number in advance. In practical situations, it is often difficult to obtain this information. Therefore, it is necessary to introduce a regularization approach to GVMD, namely the SGVMD algorithm, to achieve the sequential extraction of each modal component of a time series in the case of unknown modal numbers.

The SGVMD algorithm is not sensitive to the initial values of the modes. The output order of the components is also flexible, and adjusting the initial values can change the extraction order. The convergence speed of the algorithm is also fast. In order to reduce the number of extraction iterations and further accelerate the convergence process, the initial peak of the mode can be set to the spectral peak of the current residual mixed component. Existing decomposition algorithms are highly dependent on the modal number, but the SGVMD algorithm overcomes this difficulty, which greatly extends its application scope. The SGVMD algorithm splits the mixed mode into two parts each time. In order to achieve the extraction of the current mode, the loss function established imposes stricter constraints on the current mode.

Taking non-stationary time series decomposition as an example. Firstly, we need to consider maximizing the fidelity between the extracted components and the current remaining components. Because maximizing fidelity can be achieved by minimizing residual terms, component fidelity terms can be designed as
\begin{equation} 
		\| \hat{f}_{i-1}^r(\omega) - \hat{u}_i(\omega) \|_2 
\end{equation}
Where $\hat{f}_{i-1}^r$ is the remaining part of the mixed sequence after the $(i-1)$th extraction

In order to describe narrowband characteristics in the frequency domain, the introduction of current component feature terms is also necessary. This item can be represented as
\begin{equation} 
		\| \hat{u}_{i}(\omega) (\omega - \omega_{c,i}) \|_2 
\end{equation}

with the center frequency $w_{c,i}$ of the component $\hat{u}_{i}(\omega)$ equal to
\begin{equation} 
		\omega_{c,i}=\frac{\int^{+\infty}_0 \omega | \hat{u}_i^{s-1}(\omega) |^2 d\omega}{\int^{+\infty}_0  | \hat{u}_i^{s-1}(\omega) |^2 d\omega} 
\end{equation}

The above two constraints can only constrain the current mode. If we do not impose any restrictions on the remaining modes, we still cannot achieve the purpose of sequential extraction. Therefore, in order to limit the remaining modes to narrowband components, we should introduce a third term, which is expressed as follows:
\begin{equation} 
		\| \hat{f}_{i}^r(\omega) (\omega - \omega_{c,i}^r) \|_2 
\end{equation}

In this equation, $f_{i}^r(\omega)$ is the remaining signal, can be expressed as
\begin{equation} 
		\hat{f}_{i}^r(\omega) = \hat{f}_{i-1}^r(\omega) - \hat{u}_i(\omega) = \hat{f}(\omega) - \sum_{j=1}^{i-1}\hat{u}_{j}(\omega) 
\end{equation}

Also $\omega_{c,i}^r$ is the center frequency of $\hat{f}_i^r(\omega)$, similarly can be shown as
\begin{equation} 
		\omega_{c,i}^{r}=\frac{\int^{+\infty}_0 \omega | \hat{f}_i^{r}(\omega) |^2 d\omega}{\int^{+\infty}_0  | \hat{f}_i^{r}(\omega) |^2 d\omega} 
\end{equation}

In conclusion, when the component features of a time series can be defined by explicit feature terms, SGVMD can effectively solve its sequential decomposition problem. The loss function (2) introduces two penalty factors, $\alpha$ and $\beta$, but experiments have shown that the ratio between the two factors is what affects the separation results. Therefore, simplifying the equation by setting one of the factors to 1 reduces the number of parameters needed.

the loss function of the SGVMD algorithm can be rewritten as Equation (2):
\begin{equation} 
	\begin{aligned}
		&\mathcal{L}\left( {\hat{u}}_{i} \right) = \left\| {{\hat{f}}_{i - 1}^{r}(\omega) - {\hat{u}}_{i}(\omega)} \right\|_{2}^{2} + \alpha\left\| {{\hat{u}}_{i}(\omega)\left( {\omega - \omega_{c,i}} \right)} \right\|_{2}^{2}\\
		&+ \beta\left\| {{\hat{f}}_{i}^{r}(\omega)\left( {\omega - \omega_{c,i}^{r}} \right)} \right\|_{2}^{2}
	\end{aligned}
\end{equation}

In this equation, $\alpha$ and $\beta$ are penalty factors. In our case, their ratio $\beta / \alpha$ would actually affect the result. So we always choose $\alpha$ = 1, leave only $\beta$ tunable. Our purpose is to minimize $\mathcal{L}(\hat{u}_i)$, which is a simple convex optimization problem. Replace $\hat{f}^r_i(\omega)$ with $\hat{f}^r_{i-1}(\omega) - \hat{u}_i(\omega)$ to emerge $\hat{u}_i(\omega)$ term, then do the differential about $\hat{u}_i(\omega)$ to $\mathcal{L}(\hat{u}_i)$ and set the result to zero, we obtain the solution,
\begin{equation} 
		\hat{u}_i(\omega) = \frac{\hat{f}^r_{i-1}(\omega)(1 + \beta(\omega - \omega^r_{c,i})^2)}{1 + \alpha (\omega - \omega_{c,i})^2 + \beta(\omega - \omega^r_{c,i})^2}
\end{equation}

Since $\omega_{c,i}$ and $\omega^r_{c,i}$ are both related to $\hat{u}_i(\omega)$, we can iterate to obtain a solution starting from an initial value noted as $\hat{u}^0_i(\omega)$. T
The process of SGVMD algorithm applied to non-stationary time series decomposition is shown in Algorithm 1, and the outer and inner stop threshold $\epsilon$ and $\eta$ are defined by user.

\begin{algorithm}[H]
	\caption{SGVMD algorithm for non-stationary time series.}\label{alg:alg1}
	\begin{algorithmic}[1]
		\STATE {\textsc{INPUT:time series dataset }$Y = \{ y_1,y_2,\cdots,y_n\};$}
		\STATE {\textsc{OUTPUT:Mode components }$\hat{u_1},\hat{u_2},\cdots,\hat{u_i};$}
		\STATE \hspace{0.5cm}$ \textbf{Initialize } \epsilon , \alpha  \textbf{ and }\beta  ,\textbf{ let } \hat{f}_0^r(\omega) = \hat{f}(\omega);$
		\STATE \hspace{0.5cm}$ \textbf{Let } i = 1;$
		\STATE \hspace{0.5cm}$ \textbf{WHILE } \left\| {{\hat{f}}_{i}^{r}(\omega)} \right\|_{2}^{2} > \epsilon \textbf{ DO }$
		\STATE \hspace{0.5cm}\hspace{0.5cm}$ \textbf{Initialize } \hat{u}_i^0(\omega) \textbf{ and } \eta;$
		\STATE \hspace{0.5cm}\hspace{0.5cm}$ \textbf{Let } s = 1;$
		\STATE \hspace{0.5cm}\hspace{0.5cm}$ \textbf{DO}$
		\STATE \hspace{0.5cm}\hspace{0.5cm}\hspace{0.5cm}$ \omega_{c,i}^{s-1}=\frac{\int^{+\infty}_0 \omega | \hat{u}_i^{s-1}(\omega) |^2 d\omega}{\int^{+\infty}_0  | \hat{u}_i^{s-1}(\omega) |^2 d\omega} ;$
		\STATE \hspace{0.5cm}\hspace{0.5cm}\hspace{0.5cm}$ \hat{f}_i^{r,s-1}(\omega) = \hat{f}_{i-1}^r(\omega) - \hat{u}_i^{s-1}(\omega);$
		\STATE \hspace{0.5cm}\hspace{0.5cm}\hspace{0.5cm}$ \omega_{c,i}^{r,s-1}=\frac{\int^{+\infty}_0 \omega | \hat{f}_i^{r,s-1}(\omega) |^2 d\omega}{\int^{+\infty}_0  | \hat{f}_i^{r,s-1}(\omega) |^2 d\omega} ;$
		\STATE \hspace{0.5cm}\hspace{0.5cm}\hspace{0.5cm}$ \hat{u}_i^s(\omega) = \frac{ \hat{f}_{i-1}^r(\omega)(1+\beta (\omega - \omega_{c,i}^{r,s-1})^2) }{1 + \alpha (\omega - \omega_{c,i}^{s-1})^2 + \beta(\omega - \omega_{c,i}^{r,s-1})^2}; $
		\STATE \hspace{0.5cm}\hspace{0.5cm}\hspace{0.5cm}$ s = s + 1; $
		\STATE \hspace{0.5cm}\hspace{0.5cm}$\textbf{WHILE } \| \hat{u}_i^{s-1}(\omega) - \hat{u}_i^{s-2}(\omega) \|_2^2 > \eta $
		\STATE \hspace{0.5cm}\hspace{0.5cm}$ \hat{u}_i(\omega) = \hat{u}_i^s(\omega); $
		\STATE \hspace{0.5cm}\hspace{0.5cm}$\hat{f}_i^r(\omega) = \hat{f}_{i-1}^r(\omega) - \hat{u}_i(\omega); $
		\STATE \hspace{0.5cm}\hspace{0.5cm}$i = i + 1;$
		\STATE \hspace{0.5cm}$\textbf{END WHILE }$
	\end{algorithmic}
	\label{alg1}
\end{algorithm}

When the component features of a time series can be defined by explicit feature terms, SGVMD can effectively solve its sequential decomposition problem. The loss function equation (4) introduces two penalty factors $\alpha$ and $\beta$, but experiments have found that the ratio of the two factors affects the separation results. Therefore, simplifying it by setting one of them to 1 reduces the number of parameters.

\subsection{ARIMA model}
ARIMA, an acronym for Autoregressive Integrated Moving Average, finds its applicability in time series forecasting by addressing stationary time series data. The essence of ARIMA lies in its amalgamation of autoregressive (AR) and moving average (MA) components, coupled with a differencing operation (I) that transforms the original series into a stationary one. This amalgamation equips ARIMA to model both short-term and long-term dependencies, rendering it particularly effective for capturing complex temporal patterns inherent in diverse datasets.

The ARIMA model is defined by three key parameters: $p$, $d$, and $q$. Here, p represents the autoregressive order, $d$ denotes the differencing order, and $q$ signifies the moving average order. Mathematically, the ARIMA$(p, d, q)$ model is expressed as:
\begin{equation} 
		Y_t = \mu + \phi_1 Y_{t-1} + \cdots + \phi_p Y_{t-p} + N_t - \theta_1 N_{t-1} - \cdots  - \theta_q N_{t-q}
\end{equation}
Where $Y_t$ is the observed value at time $t$, $\mu$ is the mean of the series, $\phi_i$ represents the autoregressive coefficients, $N_t$ is the white noise error term at time $t$, $\theta_i$ denotes the moving average coefficients.

ARIMA exhibits notable strengths, such as its versatility in capturing both short-term and long-term dependencies, adaptability to various types of time series data, and the ability to provide reliable forecasts. However, its efficacy is contingent upon the assumption of linearity, which may limit its performance in scenarios involving highly non-linear data. Additionally, ARIMA may face challenges when dealing with irregularly spaced or sparse time series data, and its performance may degrade when confronted with abrupt structural changes in the underlying process. Thus, while ARIMA stands as a formidable tool in time series forecasting, its application necessitates a discerning consideration of the inherent characteristics of the data at hand.

\subsection{Sequential General Variational Mode Decomposition Method for Time Series components prediction}
Generally, the components extracted by SGVMD can be divided into trend components and AM-FM components. This prediction method is suitable for predicting the AM-FM components extracted by SGVMD, and the trend component can be directly predicted using ARIMA. The implementation process of the SGVMD-E-ARIMA is introduced as follows:(1) Extract the upper and lower envelope of each component after the decomposition by the SGVMD algorithm. Here, the Hilbert transform and local interpolation method will be used, and the result is shown in Fig. (a);(2) For each extracted envelope, build a corresponding ARIMA model for prediction, as shown in Fig. (b);(3) After obtaining the predicted results of the upper and lower envelopes, the prediction of the AM-FM component can be obtained by combining with the frequency-modulated (FM) component, as shown in Fig. 1:

\begin{figure}[htbp]
\centering
\subfigure[]{
\includegraphics[width=7.4cm]{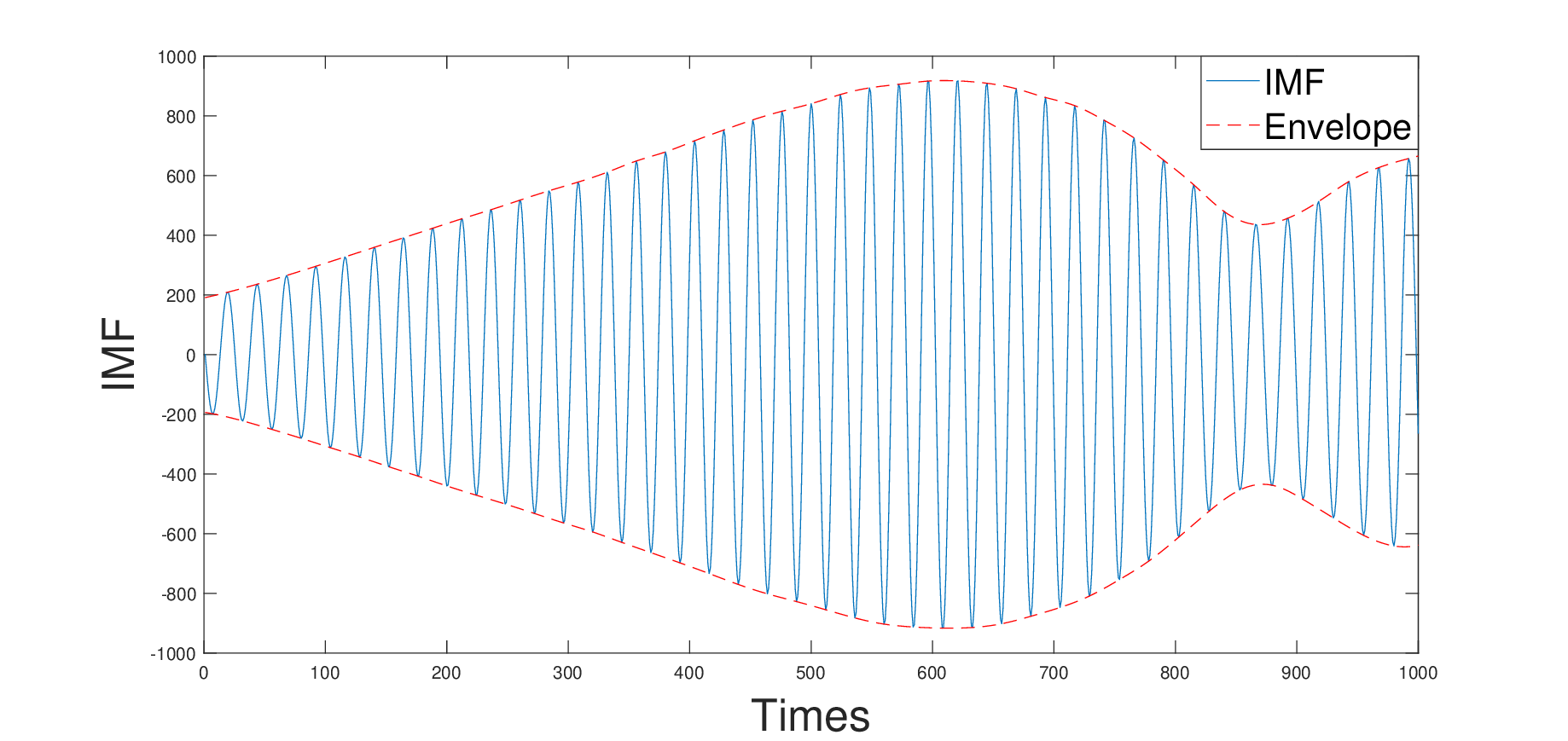}
}
\quad
\subfigure[]{
\includegraphics[width=7.4cm]{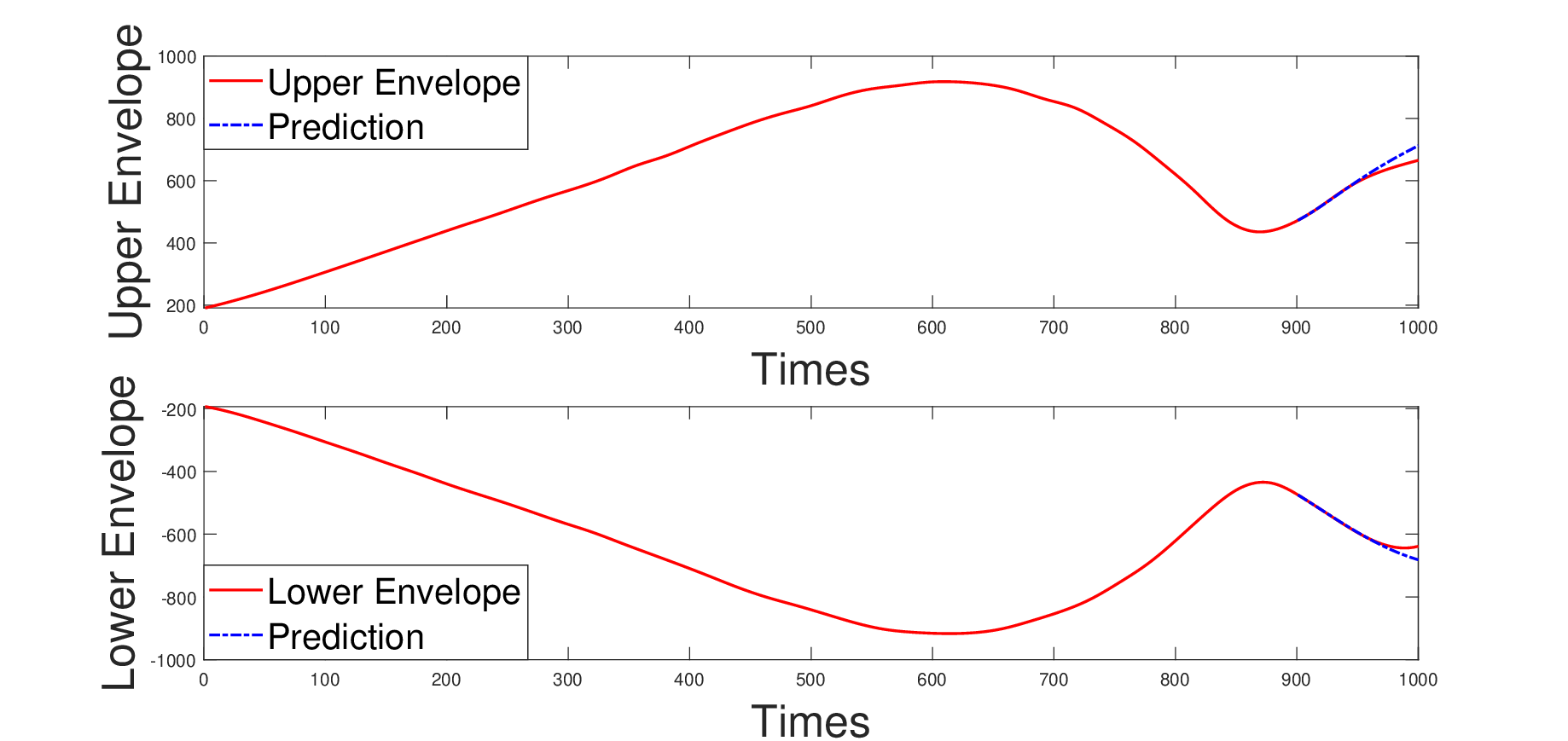}
}
\subfigure[]{
\includegraphics[width=\linewidth]{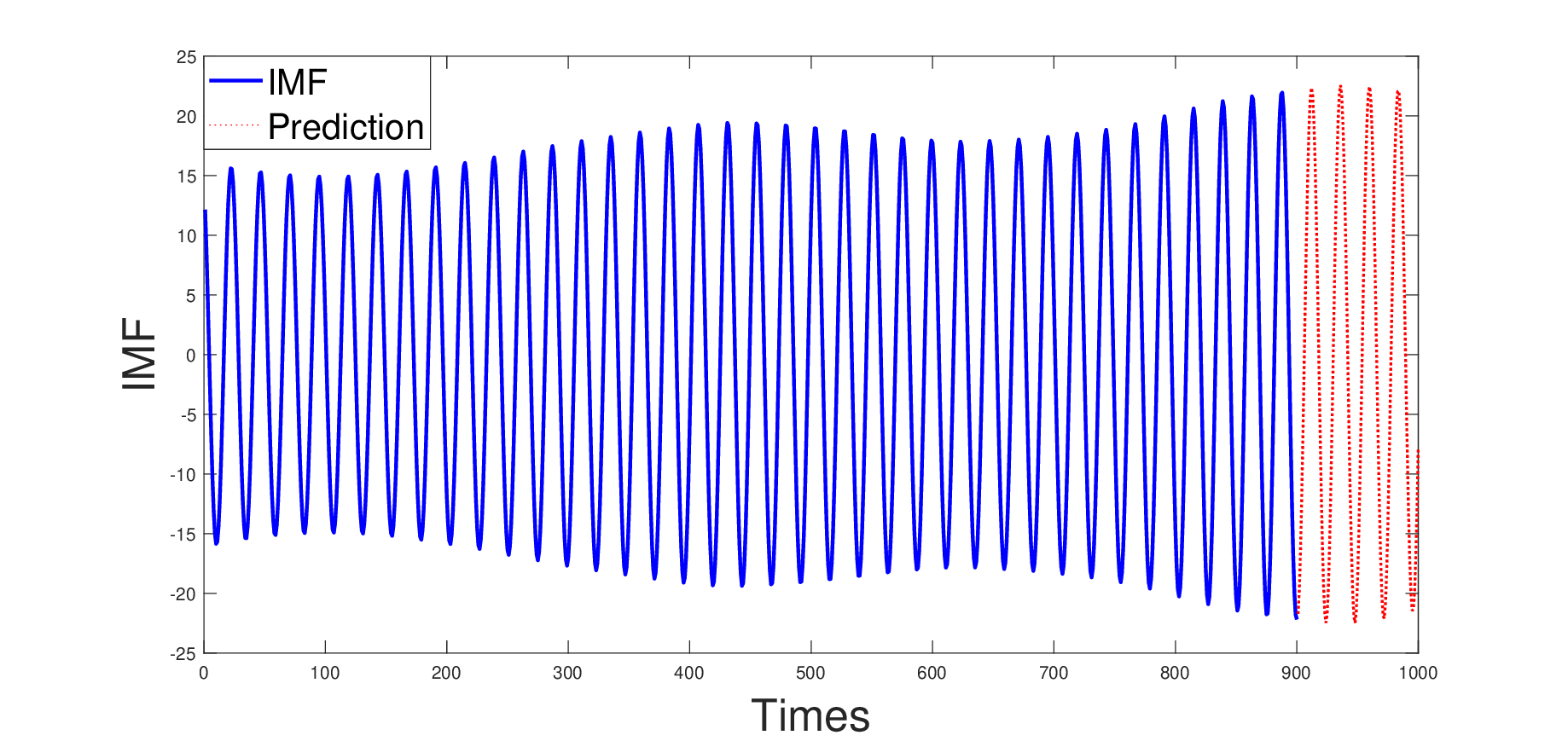}
}
\caption{Modal components and their upper and lower envelope lines and prediction results of each envelope line. And the prediction results of AM-FM components.}
\end{figure}

We propose the combination prediction method that uses SGVMD to decompose a time series, and then operate every component by predicting the AM-FM component. This process is called the SGVMD-ARIMA. In the decomposition phase, SGVMD regards the time series as a non-stationary mixed signal and uses SGVMD to decompose it into trend signals and AM-FM signals. In the prediction phase, this method distinguishes between the two types of components. For the trend signal, the ARIMA method is directly applied, while for the AM-FM component, the amplitude envelope is obtained, the envelope is predicted, and then combined with the frequency modulation component to obtain more accurate results. This process is called the SGVMD-E-ARIMA. The meaning of E in it refers to the envelope prediction process. 
These two processes are given by Fig. 2 beneath.

\begin{figure}[h]
\center
	  \includegraphics[width=\linewidth]{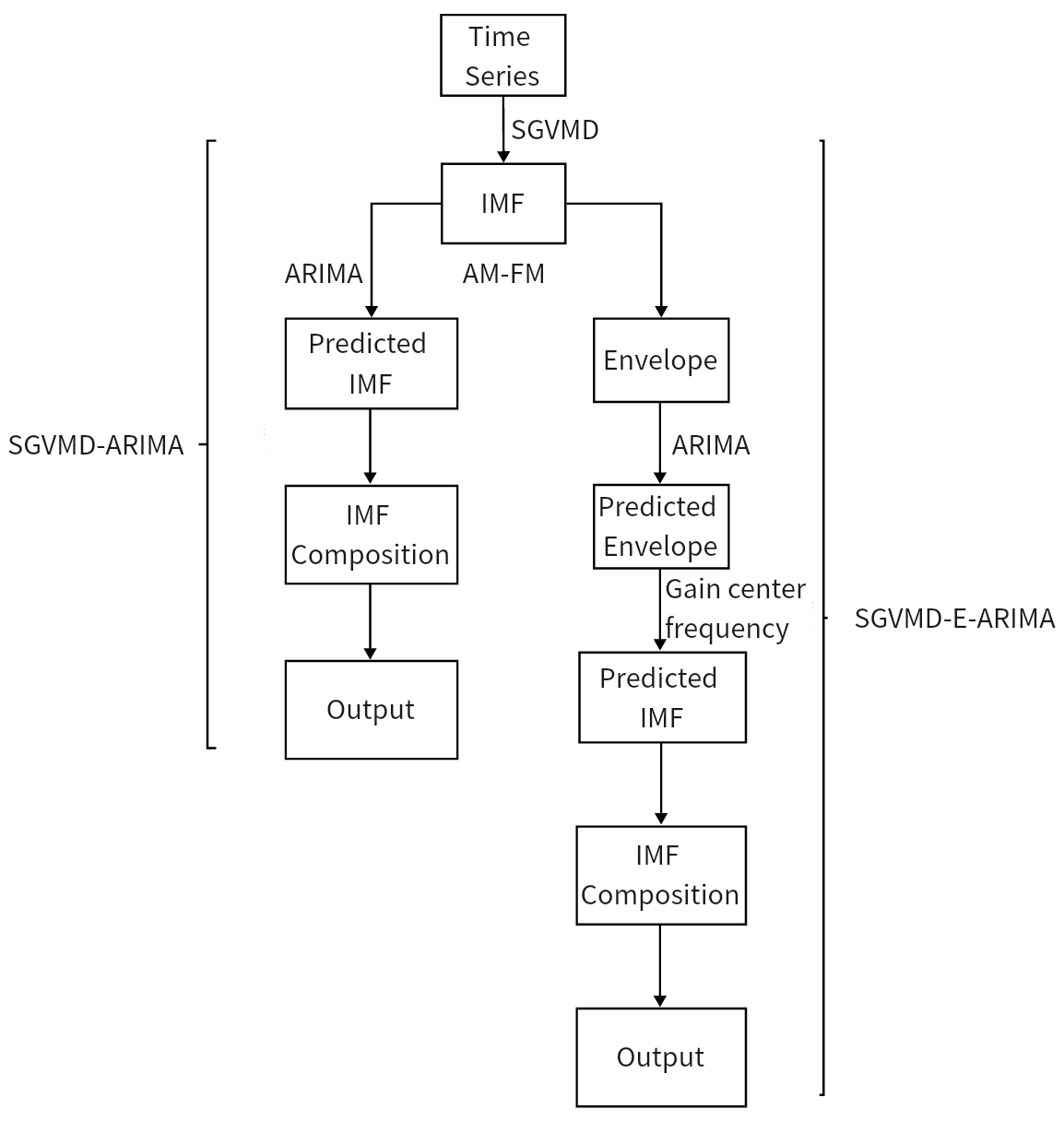}
\caption{Prediction results of AM-FM components.}
\end{figure}

\section{Experimental results and evaluation}
In this section, we will use these two aforementioned methods to predict two examples and compare their prediction performance with five other methods. These six methods can be divided into two categories: the first category is the single-model prediction method, which includes the ARIMA model, Holt-Winter model and Long Short-Term Memory (LSTM); the second category is the combination prediction method, including the EMD-ARIMA model for time series, SGVMD-ARIMA and SGVMD-E-ARIMA combination model that we proposed above.

\subsection{historical sales of an online store}
The selected time series in this example is the historical sales of a certain product in an online store, and a segment of the time series that can highlight the trend is selected as the experimental dataset. The total number of data points is 1000, and its time series plot is shown in Fig. (a):

\begin{figure}[htbp]
\centering
\subfigure[]{
\includegraphics[width=7.4cm]{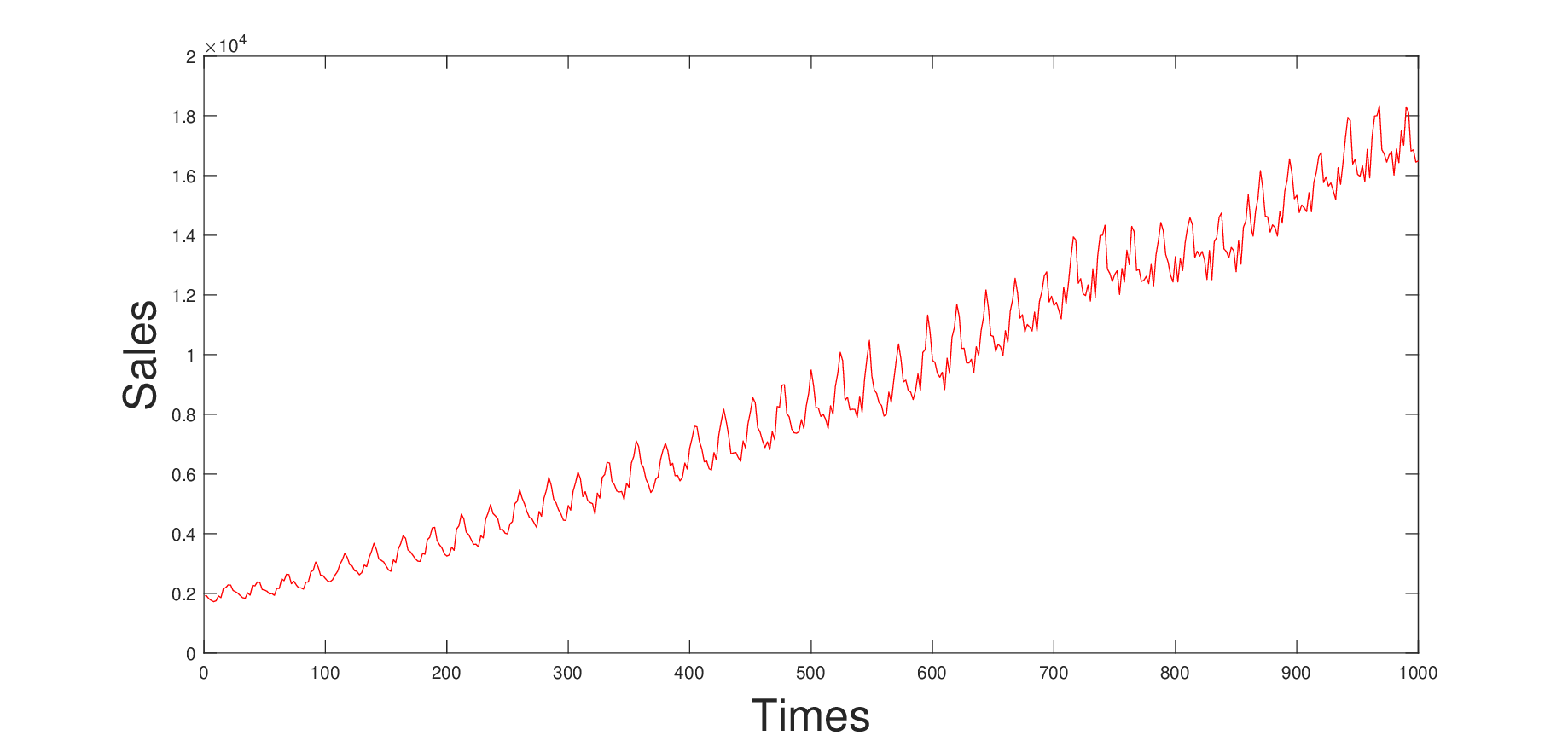}
}
\quad
\subfigure[]{
\includegraphics[width=7.4cm]{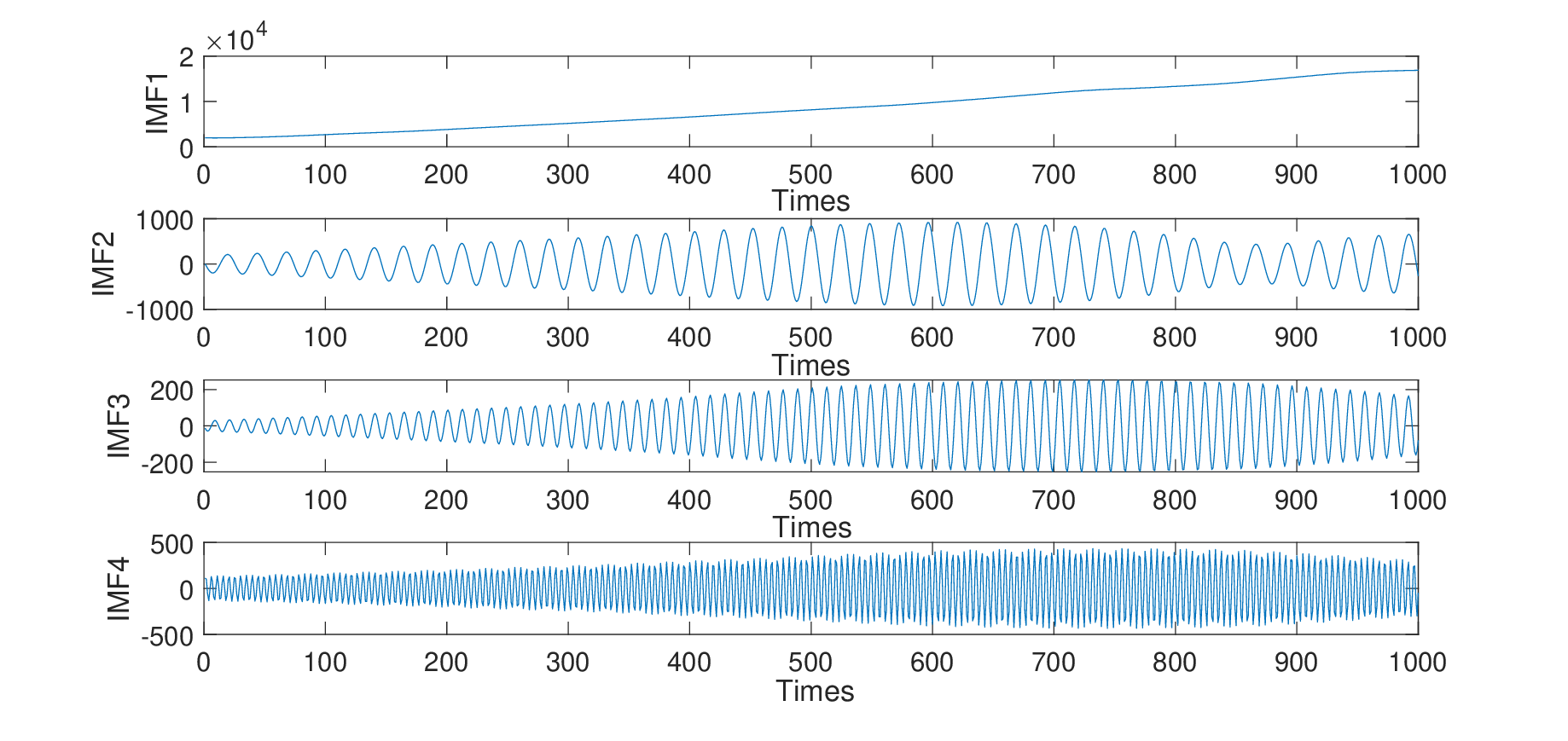}
}
\subfigure[]{
\includegraphics[width=\linewidth]{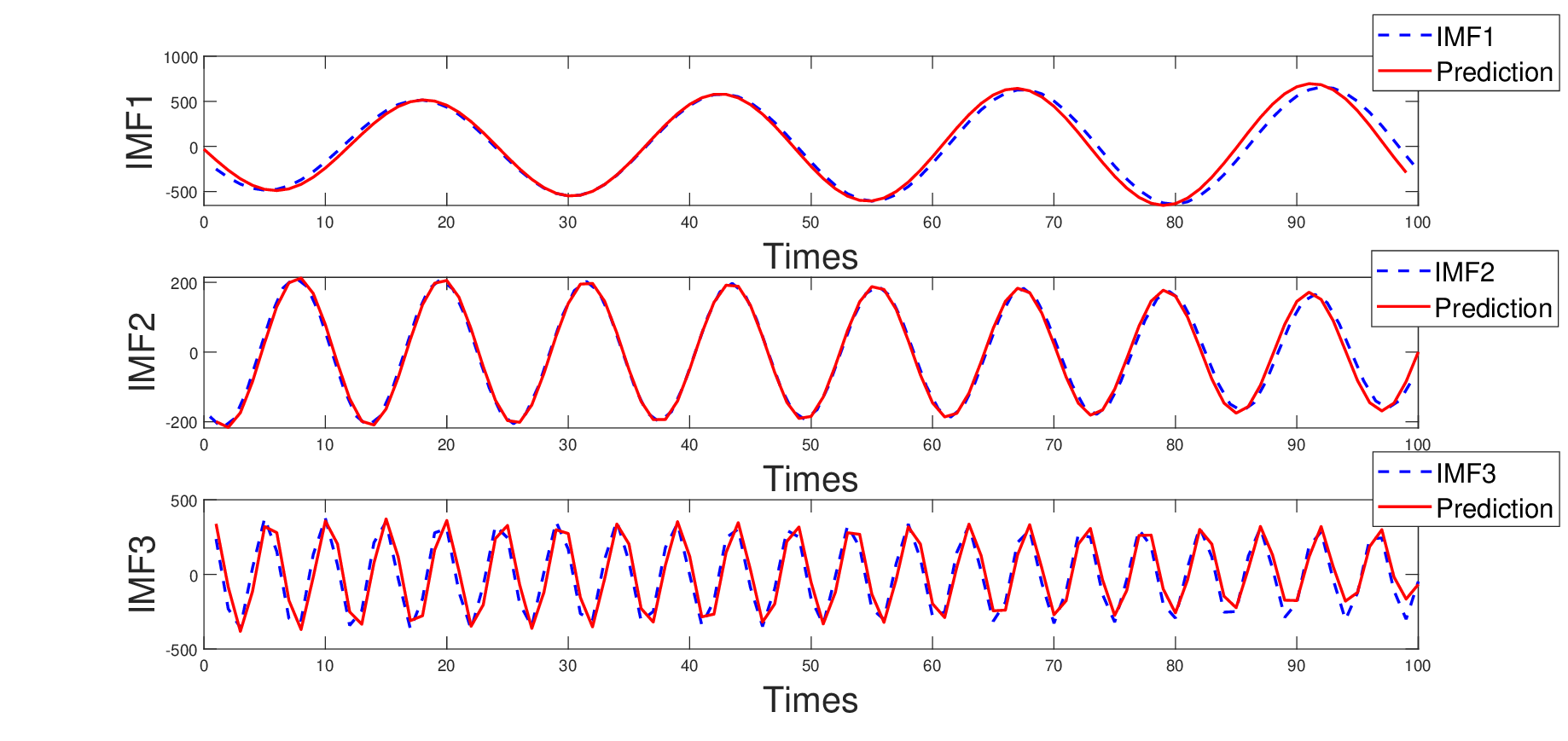}
}
\caption{Time series plot of experimental data for Example A and the results of IMF component analysis for the Example A. And Optimized prediction results for each AM-FM component.}
\end{figure}

The data with a length of 1000 is divided into a training set and a test set, with the ratio of 9:1. For the single ARIMA model method, different parameter combinations are compared to obtain the best fitting effect of the ARIMA(9,1,8) model on the original sequence. Since the seasonal variation of the time series in this example is proportional to the sequence level, the Holt-Winter method adopts a multiplicative model to model the sequence. After multiple experiments, the optimal parameter combination for predicting this time series is obtained as $\alpha$=0.91, $\beta$=0.5, and $\gamma$=1. For the LSTM, number of hidden units is 200, learn-rate-drop-period is 400, learn-rate-drop-factor is 0.15. Regarding the EMD-ARIMA method, we first decompose the original time series into 7 IMF components, including residuals, through EMD. Then, we construct corresponding ARIMA models for each of these 7 components for prediction. SGVMD-ARIMA also belongs to a combination prediction model. Firstly, the SGVMD algorithm is used to separate the various components of the original time series. Several principal components (as shown in Fig. (b) are selected, and each of these principal components is predicted using the ARIMA method. Finally, the results are combined.

The specific process of the combination model that we proposed is as follows:

Firstly, the SGVMD algorithm is applied to decompose the time series. In this process, we need to adjust the penalty factors $\alpha$ and $\beta$ continuously until the extracted modal components meet the previous assumptions. After the SGVMD decomposition of the original time series, four components can be obtained, as shown in Fig. (b). The result in Fig. (b) conforms to the previous assumption: component 1 is the tendency component, while the other three components are AM-FM components with certain amplitude and central frequency.

Secondly, in the envelope prediction process, for the AM-FM components, the upper and lower envelope lines are first obtained, and ARIMA models are constructed separately for prediction. Finally, the predicted envelope function is multiplied by the pure frequency modulation signal of the component to obtain the predicted sequence. In the process of restoring the predicted sequence, there is a "lag" phenomenon in the phase. Therefore, in order to optimize the prediction effect, the phase directly obtained from the spectrum needs to undergo some minor processing, such as rounding. The prediction results of each component after optimization are shown in Fig. (c). It can be seen from the figure that the optimized predicted sequence basically matches the original sequence, which also proves the feasibility of the new prediction method. For the trend component 1, the ARIMA algorithm is directly applied for prediction.

\subsubsection{Prediction Results}
The prediction results of the 100 data points on Sales for each model shows below. It can be seen that although there is a lag phenomenon between the predicted values of the LSTM model and the true values, the trend is consistent. Although the single ARIMA model and Holt-Winter model have no obvious lag, the error for each point is too sharp. The predicted values of the SGVMD-ARMA model are closer to the true values and have higher accuracy.

\begin{figure}[h]
\center
	\includegraphics[width=\linewidth]{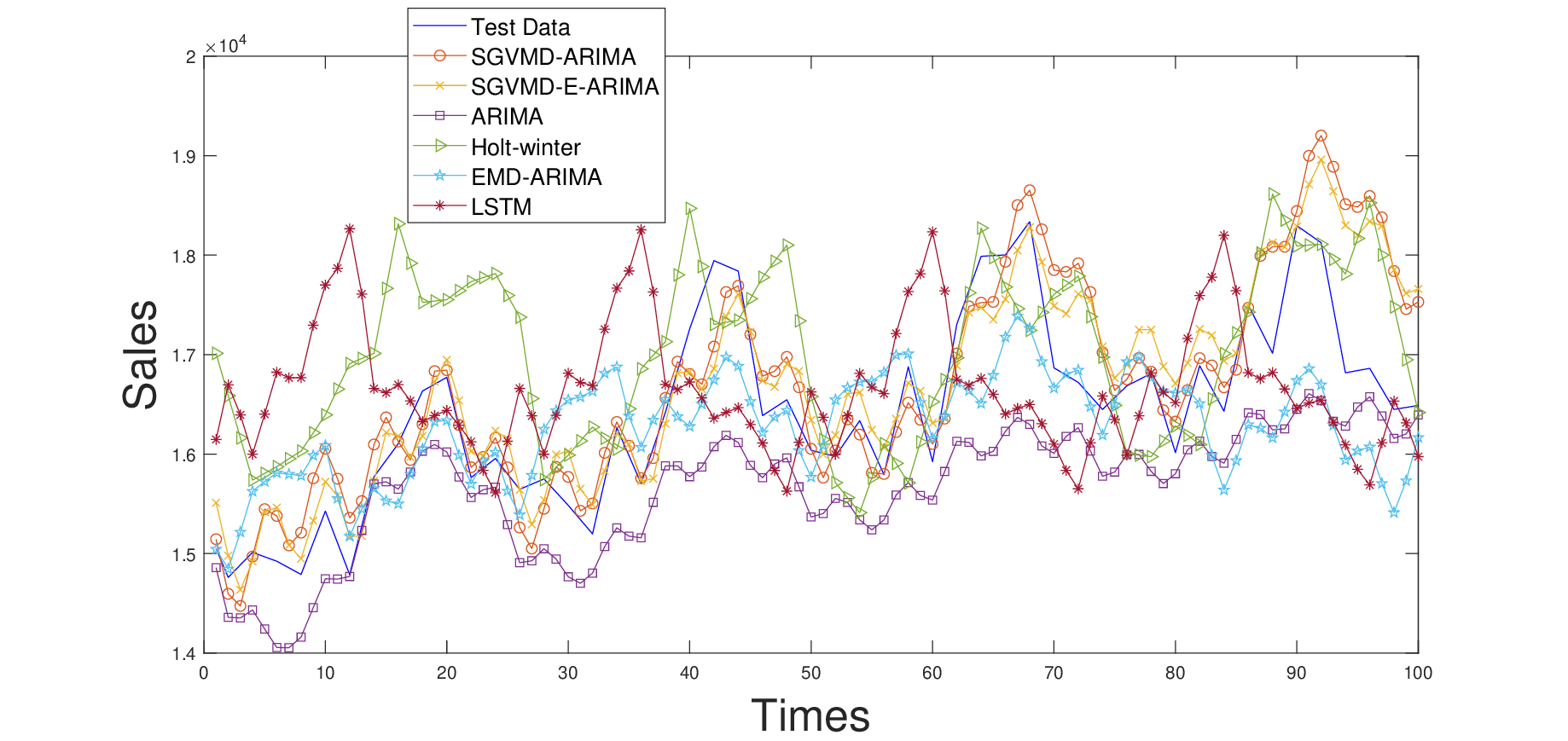}
\caption{Comparison for models in example A.}
\end{figure}

After adjusting parameters, predict future time series data with the adjusted model. During prediction, modification and optimization based on the predicted results are necessary to improve prediction accuracy and reliability. Finally, after predicting, evaluate the trained model with reserved testing dataset. Common evaluation metrics include Mean Squared Error, Root Mean Squared Error, Mean Absolute Error, and so on. Model-based time series forecasting has the advantages of better reflecting the features and regularities of the data and having high prediction accuracy and reliability. In this paper, we will use three objective evaluation metrics, namely Root Mean Squared Error (RMSE), Mean Absolute Percentage Error (MAPE), and Mean Absolute Error (MAE), to evaluate the prediction performance:

\begin{equation} 
	\begin{split}
		RMSE\left( {y,\hat{y}} \right) = {\sqrt{\frac{1}{n}{\sum_{i = 1}^{n}\left( {y - {\hat{y}}_{i}} \right)^{2}}}}
	\end{split}
\end{equation}

\begin{equation} 
	\begin{split}
		MAPE\left( {y,\hat{y}} \right) = \frac{1}{n}{\sum_{i = 1}^{n}\left|\frac{y - {\hat{y}}_{i}}{y} \right|}
	\end{split}
\end{equation}

\begin{equation} 
	\begin{split}
		MAE\left( {y,\hat{y}} \right) = \frac{1}{n}{\sum_{i = 1}^{n}\left|{y - {\hat{y}}_{i}}\right|}
	\end{split}
\end{equation}
Where $y$ is the actual value of the time series, $\hat{y}$ is the predicted value, and $n$ is the number of data points in the time series.

\begin{table*}[htbp]		
	\resizebox{\textwidth}{!}{
	\centering		
	\begin{tabular}{cccc}  
	\toprule
	Model & RMSE & MAPE & MAE\\
	\midrule
	ARIMA & 920.539008 & 0.046712 & 784.678901  \\
	Holt-Winter & 1003.134259 & 0.051554 & 831.312812  \\
	LSTM & 775.315002 & 0.035625 & 573.177551  \\
	EMD-ARIMA & 717.039878 & 0.035505 & 589.845088  \\
	SGVMD-ARIMA & 589.603168 & \textbf{ 0.025120} & \textbf{416.3842711} \\
	SGVMD-E-ARIMA & \textbf{560.219180} & 0.025635 & 424.867215 \\
	\bottomrule
	\end{tabular}
			}
\caption{Comparison for results of prediction models.}
\end{table*}

\subsubsection{Evaluation and Analysis of Prediction Results} Fig. 4 shows the comparison results of prediction methods. The following conclusions can be drawn from the result:
(1) Directly constructing ARIMA models for the original time series has certain reference value, but this reference value is very limited. As time progresses, the error between the predicted values and the actual values continues to increase, and the details are lost severely.
(2) When using the Holt-Winter method for prediction, the trend information and seasonal information of the original time series are relatively well preserved. However, this method also has an obvious shortcoming, which is that the details are lost severely.
(3) For the LSTM, if the training set not suitable enough, it will make it difficult for the model to learn long-term dependencies, thereby affecting prediction accuracy and easily causing overfitting. The model's prediction results are often difficult to explain, which may not be suitable for some application scenarios like the prediction of sales that require interpretability.
(4) The basic trend of the prediction results using the EMD-ARIMA combination method is similar to the original time series, but with errors in the amplitude of the changes. This is because the EMD algorithm decomposes the high-frequency and stable components too early, leaving the residual to rely on too much weights for representing the trend.
(5) As a combination model, the SGVMD-ARIMA and the SGVMD-E-ARIMA are both better than single prediction models and the EMD-ARIMA combination method in terms of prediction accuracy, trend restoration, and improvement of detailed features. However, it cannot be judged subjectively from a visual perspective which one is better, and this issue can be clearly solved in objective evaluations above.

\subsection{The Production of Australian Beer}
This experiment used Australian beer production as the experimental dataset to investigate the performance of the EMD-based prediction algorithm in forecasting non-monotonic trend and non-seasonal time series. Fig. 5 shows the time series plot of the dataset. The experimental process of the prediction methods is described below.

\begin{figure}[htbp]
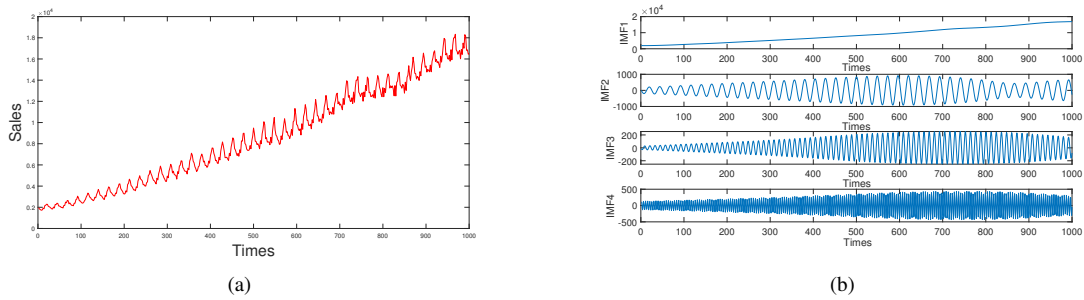

\centering
\subfigure[]{
\includegraphics[width=7.4cm]{exp1_1}
}
\quad
\subfigure[]{
\includegraphics[width=7.4cm]{exp1_2}
}

\caption{The time series plot of Australian beer production with the results of principal IMF analysis for this Example.}
\end{figure}

For the single ARIMA model method, different parameter combinations are compared to obtain the best fitting effect of the ARIMA(4,1,3) model on the original sequence. For the Holt-Winter, the optimal parameter combination for predicting this time series is obtained as $\alpha$=0.5, $\beta$=0.5, and $\gamma$=1. When training with LSTM, number of hidden units is 400, learn-rate-drop-period is 400, learn-rate-drop-factor is 0.15. Regarding the EMD-ARIMA method, we first decompose the original time series into 9 IMF components, including residuals, through EMD. The SGVMD algorithm is used to separate the various components of the original time series. Five representative principal components are selected to show in Fig. 6, and each of these principal components is predicted using the ARIMA method. Finally, the results are combined.

\subsubsection{Prediction Results}
The comparative results of prediction performance obtained from every prediction method are shown in Fig. 6.

\begin{figure}[h]
\center
	\includegraphics[width=\linewidth]{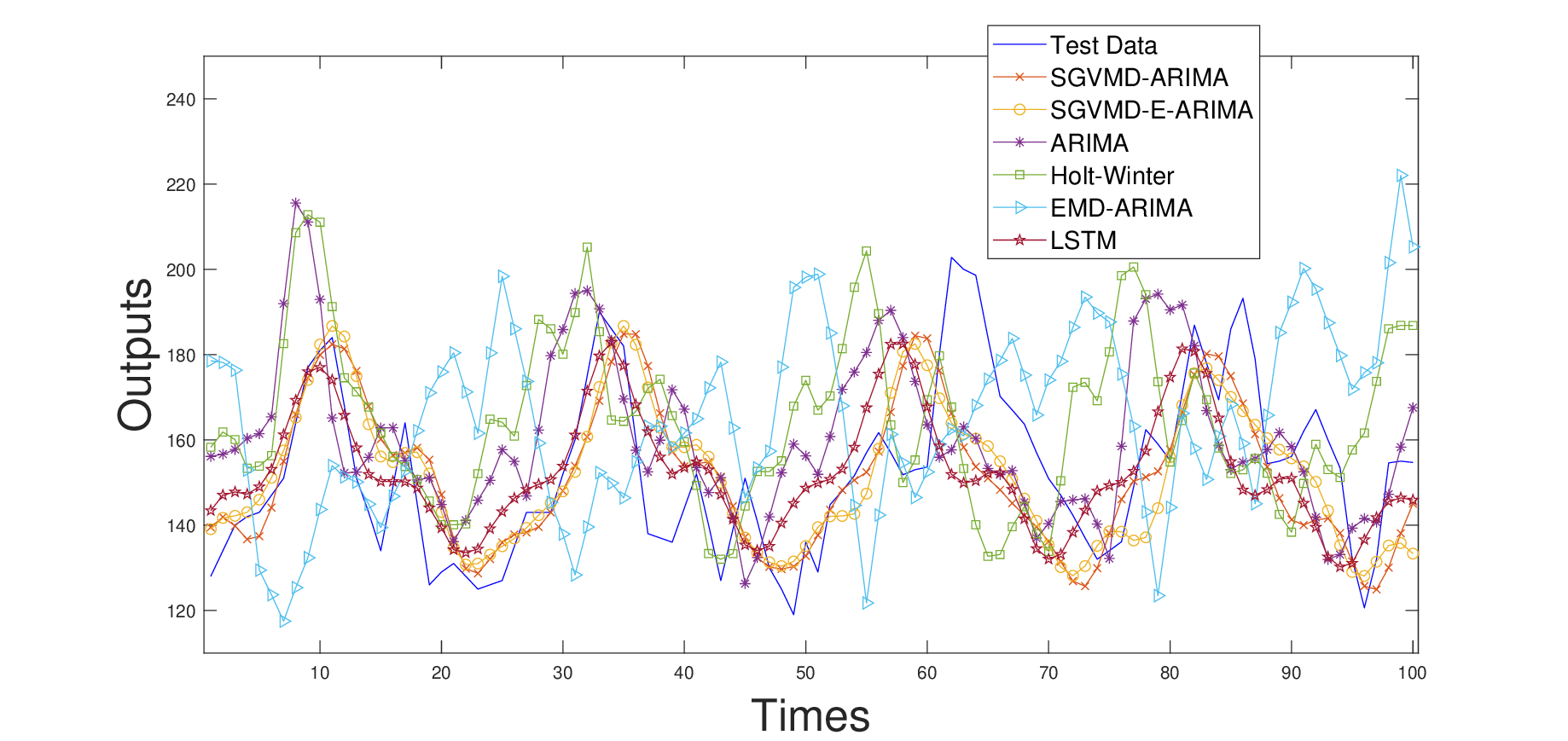}
\caption{Comparison for models in example B.}
\end{figure}

From a subjective perspective, the methods of single models such as Holt-Winter, ARIMA, and LSTM have relatively poor prediction performance. Although they have effectively restored the trend and cyclic information of time series, they all have different phase lags in various situations. The prediction performance of the three combination prediction methods is not significantly different, and their specific advantages and disadvantages can be obtained by calculating objective indicators.

 \begin{table}[htbp]	
	\resizebox{\textwidth}{!}{	
 	\centering		
 	\begin{tabular}{cccc}  
 		\toprule
 		Model & RMSE & MAPE & MAE\\
 		\midrule
 		ARIMA & 27.986811 & 0.158452 & 23.626623  \\
 		Holt-Winter & 34.498434 & 0.201383 & 29.460192  \\
 		LSTM & 22.555809 & 0.128462 & 19.437229  \\
 		EMD-ARIMA & 17.007021 & 0.085877 & 13.264099  \\
 		SGVMD-ARIMA & \textbf{15.105131} & \textbf{0.076254} & \textbf{11.781843}  \\
 		SGVMD-E-ARIMA & 16.082900 & 0.078981 & 12.210334 \\
 		\bottomrule
 	\end{tabular}
  				}
\caption{Comparison for results of  prediction models.}
 \end{table}
 
Table 2 presents the RMSE, MAPE, and MAE values for prediction outputs applied to the Example B dataset. 

\subsubsection{Evaluation and Analysis of Prediction Results}The following information can be inferred from the table: (1) The SGVMD-ARIMA combination prediction method is the best-performing method in terms of evaluation indicators, while the Holt-Winter method has the least desirable prediction effect. (2) Although LSTM is a machine learning algorithm, the prediction effect of a single prediction method is still inferior to that of a combination prediction model. (3) From the perspective of sequence decomposition, the SGVMD algorithm better preserves sequence information compared to the EMD algorithm. (4) From the perspective of component prediction, directly predicting the time sequence by constructing an ARIMA model is more effective. There are two reasons for this: firstly, the problem of phase "lag" was not well resolved in this instance; secondly, the prediction performance of the ARIMA model for the envelope line was not satisfactory, leading to errors in the amplitude of the predicted sequence.

\section{Conclusion And Prospect}

In this paper, a new method of time series prediction, the SGVMD-ARIMA and SGVMD-E-ARIMA is proposed from the perspective of signal processing. In terms of time series decomposition, a new signal separation algorithm SGVMD is proposed. It can decompose the sequence in the second order under the premise of unknown number of modes, and extract a series of modal components with narrowband characteristics in the frequency domain. In terms of prediction, a prediction method for AM-FM component is proposed, that is, using ARIMA to predict the envelope. By comparing ARIMA model, Holt-Winter model, LSTM, EMD-ARIMA combination forecasting method, it is found that the new algorithm SGVMD-ARIMA and SGVMD-E-ARIMA have the characteristics of high feasibility and accuracy in predicting highly non-stationary time series. In the experiment carried out in this paper, compared with single model , its accuracy can be improved by nearly 40$\%$.

Based on the research in this paper, the SGVMD-ARIMA and the SGVMD-E-ARIMA will have broad application prospects in the prediction of non-stationary time series, especially in the field of economic analysis. Because the economic sequence is easily affected by external factors such as culture and society, it is highly non-stationary and tends to show a certain trend as time progresses. However, this method still has some shortcomings, and future improvement and improvement work will mainly focus on the following aspects: (1) Setting of penalty factors. Although SGVMD algorithm considers the composition of loss function more comprehensively, it realizes the sequential separation of components. Although it has a relatively wide range of allowable values of penalty factors, it still has the problem that the setting of penalty factors depends on experience. Therefore, the de-experienced research of penalty factors is the focus of the next step. (2) Construction of characteristic items. This is the key factor for the success of SGVMD algorithm. In the new algorithm proposed in this paper, for non-stationary time series, "narrow band" is added to the loss function as a prior feature of the component. But not all time series components can meet this assumption. Different practical problems are needed to flexibly adjust feature items to balance particularity and generality. (3) Improve the problem of phase lag. In the experiment, using the new synthetic prediction method to predict the AM-FM component will cause the phase "lag". The reason is that the unavoidable multi frequency components in the calculation of component center frequency cannot be processed, so the recovered components are not pure single frequency components, but complex frequency components with a certain change trend. The center frequency used to recover the envelope is only a certain basic frequency, so the separation parameter requirements will be more stringent. The method used in this paper is relatively crude and needs to be further analyzed and improved.



\end{document}